\documentclass[aps,prl,twocolumn,superscriptaddress,english]{revtex4}
\usepackage{amssymb}
\usepackage{graphicx}
\usepackage{amsmath}
\usepackage{soul}
\usepackage{amsthm}
\usepackage{amsfonts}
\usepackage[T1]{fontenc}
\usepackage[latin9]{inputenc}
\usepackage{array}
\usepackage{multirow}
\usepackage{color}
\usepackage{esint}
\usepackage{bm}
\usepackage{color}
\usepackage{bbm}
\usepackage{hyperref}
\usepackage{babel}
\usepackage{titlesec}
\usepackage{makecell}
\begin{document}
\global\long\def\id{\mathbbm{1}}
\global\long\def\ui{\mathbbm{i}}
\global\long\def\ud{\mathrm{d}}

 \title{Unified Symmetry Classification of Many-Body Localized Phases}  
 
 \author{Yucheng Wang}
 \thanks{Corresponding author: wangyc3@sustech.edu.cn}
 \affiliation{Shenzhen Institute for Quantum Science and Engineering, Southern University of Science and Technology,
 	Shenzhen 518055, China}
 
 \begin{abstract}
 	The Altland--Zirnbauer (AZ) tenfold way classifies disordered non-interacting systems by symmetry, but a unified symmetry classification of interacting many-body localized (MBL) phases has been lacking. Here we develop such a classification directly from the emergent local integrals of motion (LIOMs). Our scheme rests on two complementary criteria: whether a symmetry is compatible with a complete LIOM algebra, and whether it enforces many-body degeneracies. These criteria partition MBL phases into stable, fragile, symmetry-protected topological, and unstable classes, and unify AZ classes, onsite Abelian symmetries, and continuous non-Abelian symmetries within a single operator-algebraic framework. Our results provide a systematic foundation for understanding how symmetry constrains the stability and structure of MBL phases.
 \end{abstract}
 \maketitle
 
 {\em Introduction.--} Disorder-induced localization is one of the central phenomena in quantum physics.
 For non-interacting systems, Anderson localization demonstrates that coherent quantum interference can completely suppress wave transport in the presence of sufficiently strong disorder~\cite{AL1,AL2,AL3,AL4,AL5}. A major conceptual advance in this context was the Altland-Zirnbauer (AZ) symmetry classification~\cite{AZ1997,AZ1996}, which systematically classifies Hamiltonians according to their discrete symmetries and provides a unified framework for understanding localization properties across different symmetry classes~\cite{AL4,AZ1997,AZ1996,ALS1,ALS2,ALS3,ALS4,ALS5,ALS6,ALS7,ALS8,ALS9,ALS10,ALS11}. Beyond localization, the AZ scheme later became the foundation for the symmetry classification of topological insulators and superconductors~\cite{TS1,TS2}, illustrating the broad impact of symmetry-based classifications in condensed-matter physics.
 
 In interacting quantum systems, the discovery of many-body localization (MBL) revealed that sufficiently strong disorder can prevent thermalization even in the presence of interactions~\cite{MBL0,MBL1,MBL2,MBL3,MBL4}. Over the past two decades, MBL has attracted extensive theoretical and experimental attention because of its profound implications for quantum transport, thermalization, quantum information storage, and non-equilibrium dynamics~\cite{MBL0,MBL1,MBL2,MBL3,MBL4,MBL5,MBL6,MBL7,MBL8,MBL9,MBL10,MBL11,MBL12,MBL13,MBL14,MBL15,MBL16,MBL17,MBL18}. 
 Existing results on symmetry in MBL have largely been obtained on a case-by-case basis.
 Abelian symmetries, such as U(1) particle-number conservation, are known to be compatible with stable MBL in one dimension \cite{classA1,classA2,classA3}, whereas continuous non-Abelian symmetries, most notably SU(2), generically preclude localization and destabilize the MBL phase \cite{SU21,SU22,SU23,SU24,SU25,SU26,SU27,SU28,SU29}.
 Discrete symmetries such as $\mathbb{Z}_2$ are also compatible with MBL and can support distinct localized phases, including symmetry-breaking and symmetry-protected topological (SPT) MBL phases with robust edge modes \cite{MBL2,IOMclassA,IOMSPT,SPT2,SPT3,SPT4,SPT5,SPT6,SPTMBL2}. These works establish important symmetry constraints for individual symmetry classes, but a general classification principle applicable across symmetry settings has remained absent. To our knowledge, no unified framework for MBL, analogous to the role played by the AZ classification for non-interacting systems, has yet been developed.
 
 In this work, we fill this gap by developing a unified symmetry classification of MBL phases formulated directly at the level of local integrals of motion (LIOMs)~\cite{MBL1,IOMclassA,IOMSPT}. Within this framework, the stability of MBL under a given symmetry is determined by two criteria: whether the symmetry admits a complete LIOM algebra, and whether it enforces degeneracies. These criteria classify MBL phases into four classes--stable, fragile, SPT, and unstable--and unify AZ, onsite Abelian, and continuous non-Abelian symmetries under a single operator-algebraic framework. 

 {\em Symmetry Constraints on Many-Body Localization.--}
 We consider a generic interacting lattice system with a finite-dimensional
 local Hilbert space on each site $i$.
 The operators $\sigma_i^\alpha$ ($\alpha = x,y,z$) denote a set of microscopically local observables, such as Pauli matrices acting on a
 physical spin-$1/2$ degree of freedom or, more generally, an on-site
 operator basis. These bare operators do not commute with the interacting Hamiltonian and
 are therefore not conserved.
 In a MBL phase, there exists a quasi-local unitary
 transformation $U$ that maps the microscopic operators to dressed, emergent
 conserved operators, $\tau_i^\alpha = U \sigma_i^\alpha U^\dagger$.
 Among them, the operators $\tau_i^z$ form a complete set of independent
 LIOMs~\cite{MBL1,IOMclassA,IOMSPT} satisfying $[\tau_i^z, H] = 0$ for every site $i$.
 Consequently, the physical Hamiltonian can be written directly in terms of these LIOMs as
 \begin{equation} \label{eq:H_lbit}
 	H
 	= \sum_i h_i \, \tau_i^z
 	+ \sum_{i<j} J_{ij} \, \tau_i^z \tau_j^z
 	+ \sum_{i<j<k} K_{ijk} \, \tau_i^z \tau_j^z \tau_k^z
 	+ \cdots.
 \end{equation}
 where $h_i$ are renormalized local fields. Equation~(\ref{eq:H_lbit}) is the standard phenomenological description of a fully MBL phase in terms of LIOMs.
 The effective couplings are quasi-local and decay exponentially with the spatial extent of their operator support. Specifically,
 $|J_{ij}|
 \propto
 J_0 e^{-|i-j|/\kappa}$,
 and
 $|K_{ijk}|
 \propto
 K_0 e^{-|i-k|/\kappa}$ for $i<j<k$~\cite{MBL1},
 where the decay is governed by the full spatial span of the operator support. Here $\kappa$ denotes the localization length associated with the quasi-local unitary transformation.
 
 
 Global symmetries act on the microscopic degrees of freedom and are carried over to the emergent LIOM algebra through the quasi-local unitary transformation that underlies the MBL phase. 
 A symmetry group $G$ is compatible with the LIOM
 framework if there exists such a transformation that maps the microscopic operators to a complete set of mutually commuting quasi-local LIOMs
 $\{\tau_i^z\}$, with the Hamiltonian diagonal in this basis and
 the adjoint action of every $g\in G$ preserving the LIOM algebra. In particular, each LIOM transforms as
 \begin{equation}
 g\,\tau_i^z\,g^{-1}
 =
 f_i\bigl(\{\tau_j^z\}\bigr),
 \end{equation}
 where
 $f_i(\{\tau_j^z\})$ is a quasi-local function of the LIOMs whose
 dependence on $\tau_j^z$ decays exponentially with the distance between
 $i$ and $j$.
 
 
 The symmetry classification developed below is formulated by organizing
 different symmetry classes according to their implications for the
 stability of MBL.
 The stability is characterized by two complementary considerations.
 The first is whether the symmetry is compatible with a
 complete LIOM algebra. The second is whether, once such a LIOM description exists, the symmetry
 enforces degeneracies.
 
 We first discuss Abelian symmetries,  
 which are generally compatible with MBL. Consider a U(1) symmetry generated by a conserved charge $Q$. Since all irreducible representations of U(1) are one-dimensional, the LIOMs can be chosen to carry definite U(1) quantum numbers and commute with $Q$: 
 \begin{equation}
 	[Q,\tau_i^z]=0, \qquad e^{i\theta Q}\tau_i^z e^{-i\theta Q}=\tau_i^z.
 \end{equation}
 Thus, the U(1) symmetry preserves the LIOM algebra without mixing different LIOMs or introducing degeneracies. Discrete Abelian symmetries behave similarly: they preserve the commuting LIOM algebra, though they may act nontrivially, e.g., $g\tau_i^z g^{-1} = -\tau_i^z$.
 Since all irreducible representations of Abelian groups are
 one-dimensional, such operations never force local operators into non-commuting multiplets.
 Abelian symmetries therefore support both symmetry-breaking~\cite{IOMSPT,SPT2} and SPT MBL phases; the latter can host robust edge modes throughout the many-body spectrum, even at high energy densities~\cite{SPT2,SPT3,SPT4}.
 
 In contrast, continuous non-Abelian symmetries fundamentally obstruct stable MBL. Unlike Abelian groups, they have higher-dimensional irreducible representations, so local operators must form symmetry multiplets rather than remain independent scalars. To see the obstruction, suppose a complete LIOM algebra compatible with SU(2) symmetry existed.
 The corresponding operators would then have to transform as a vector under SU(2): $U(R)\tau_i^\alpha U^\dagger(R)=R_{\alpha\beta}\tau_i^\beta$,
 where $U(R)$ is the unitary representation of the rotation and $R_{\alpha\beta}$ is the corresponding SO(3) matrix. For example, under a rotation about the $y$ axis generated by the global spin operator $S_y=\sum_i S_i^y$, one obtains (see the Supplementary Material~\cite{SM} for details)
 \begin{equation}
 	e^{i\theta S_y}\tau_i^z e^{-i\theta S_y}
 	=
 	\cos\theta\,\tau_i^z
 	- \sin\theta\,\tau_i^x.
 	\label{eq:SU2rotation}
 \end{equation}
 Thus, no single component is invariant; rather, the symmetry rotates different components of each local multiplet into one another. The eigenvalue of $\tau_i^z$ alone therefore no longer defines a complete local conserved quantity.
 The local operators form a non-commuting $\mathfrak{su}(2)$ Lie algebra, not a commuting LIOM algebra. 
 This multiplet structure also enforces extensive degeneracies: specifying the LIOM eigenvalues does not uniquely determine an eigenstate, but leaves an internal multiplet at each site. When this holds throughout the system, the spectrum contains an extensive manifold of locally degenerate states. These degeneracies enhance resonant hybridization and facilitate thermal avalanches~\cite{SU21,SU22,SU23,SU24,SU25,SU26,SU27,SU28,SU29}, thereby destabilizing MBL. Here we emphasize the underlying
 algebraic origin: continuous non-Abelian symmetries are
 incompatible with a commuting LIOM algebra.
 
 \begin{table*}[t]
 \centering
 \renewcommand{\arraystretch}{1.25}
 \begin{tabular}{c|c|c|c|c|c}
 	\hline\hline
 	Class 
 	& $T^2$ 
 	& $C^2$ 
 	& Onsite symmetry 
 	& Constraint on LIOM algebra
 	& Classification (1D) \\
 	\hline
 	A 
 	& 0 & 0 
 	& None 
 	& No additional constraint
 	& Stable MBL \\
 	
 	AI 
 	& +1 & 0 
 	& None 
 	& Real LIOM algebra
 	& Stable MBL \\
 	
 	AII 
 	& $-1$ & 0 
 	& None 
 	& LIOM algebra with symmetry-enforced degeneracy
 	& Fragile MBL \\
 	
 	AIII 
 	& 0 & 0 
 	& None (AZ chiral $S$) 
 	& Chiral pairing
 	& Stable MBL \\
 	
 	\hline
 	D 
 	& 0 & +1 
 	& None 
 	& BdG-type LIOM algebra 
 	& Stable MBL \\
 	
 	C 
 	& 0 & $-1$ 
 	& None 
 	& BdG-type LIOM algebra 
 	& Stable MBL \\
 	
 	\hline
 	A + U(1) 
 	& 0 & 0 
 	& U(1) charge 
 	& Charge-conserving LIOM algebra
 	& Stable MBL \\
 	
 	AI + U(1) 
 	& +1 & 0 
 	& U(1) charge 
 	& Real, charge-conserving LIOM algebra
 	& Stable MBL \\
 	
 	\hline
 	A + $\mathbb{Z}_2$ 
 	& 0 & 0 
 	& $\mathbb{Z}_2$ 
 	& Linear \emph{or} projective LIOM algebra 
 	& Trivial MBL \emph{or} SPT-MBL \\
 	
 	AI + $\mathbb{Z}_2$ 
 	& +1 & 0 
 	& $\mathbb{Z}_2$ 
 	& Real linear \emph{or} projective LIOM algebra  
 	& Trivial MBL \emph{or} SPT-MBL \\
 	
 	\hline
 	A + $\mathbb{Z}_2 \times \mathbb{Z}_2$ 
 	& 0 & 0 
 	& $\mathbb{Z}_2 \times \mathbb{Z}_2$ 
 	& Projective LIOM algebra 
 	& SPT-MBL \\
 	
 	AI + $\mathbb{Z}_2 \rtimes T$ 
 	& +1 & 0 
 	& $\mathbb{Z}_2 \rtimes T$ 
 	& Real projective LIOM algebra 
 	& SPT-MBL \\
 	
 	\hline
 	Non-Abelian 
 	& -- & -- 
 	& SU(2), SU(N) 
 	& Non-commuting Lie algebra 
 	& Unstable MBL \\
 	
 	\hline\hline
 \end{tabular}
 \caption{
 	Symmetry classification of one-dimensional MBL phases
 	The first three columns specify the AZ symmetry class
 	through time-reversal ($T$) and particle-hole ($C$) symmetries.
 	The column ``Onsite symmetry'' lists additional global onsite symmetries beyond the AZ classification.
 	The column ``Constraint on LIOM algebra'' summarizes how the imposed symmetries
 	constrain the algebraic structure of the emergent LIOMs.
 	The final column, ``Classification (1D),''
 	gives the resulting symmetry classification in one dimension,
 	distinguishing stable, fragile, SPT, and unstable MBL
 	phases.
 }
 \label{tbl:MBL_full}
 \end{table*}

 Anti-unitary symmetries behave differently. Consider time-reversal symmetry $T$, which acts on the LIOMs as
 $T\,\tau_i^z\,T^{-1}
 =
 \eta_T\tau_i^z$,
 with $\eta_T=\pm1$.
 Unlike continuous non-Abelian symmetries, this transformation does not
 mix different LIOMs.
 Consequently, the complete commuting LIOM algebra remains well defined,
 and every many-body eigenstate may still be written as
 $|\Psi\rangle
 =
 |\tau_1^z,\tau_2^z,\cdots,\tau_N^z\rangle$.
 The tensor-product structure of the Hilbert space is therefore
 preserved. The essential distinction from ordinary stable MBL phases is that
 $\mathcal T^2=-1$ enforces Kramers degeneracy.
 Although this degeneracy does not invalidate the LIOM algebra, it
 introduces exact resonances between symmetry-related 
 configurations. Such resonances may ultimately destabilize the MBL phase
 on sufficiently large length and time scales~\cite{SU22}
 (see the Supplementary Material~\cite{SM}).
 Thus, while these systems can still exhibit stable MBL on
 finite length and time scales, they are expected to become unstable
 asymptotically. We therefore classify them as fragile MBL: the LIOM algebra
 remains well defined, while symmetry-enforced degeneracies are
 present.
 The other AZ symmetries, namely particle-hole symmetry
 and chiral symmetry, do not in general enforce 
 degeneracies. They therefore remain
 compatible with stable MBL.


 {\em Symmetry Classification of MBL Phases.--}
 We now present a systematic classification of MBL phases, summarized in Table~\ref{tbl:MBL_full}, which incorporates the AZ classes together with onsite Abelian and continuous non-Abelian symmetries.   
 We note that several AZ classes that are distinct in non-interacting systems,
 including BDI, DIII, CI, and CII, do not give rise to independent entries
 in the present MBL classification. The reason is that our classification is determined by the constraints a symmetry imposes on the emergent LIOM algebra, together with whether it enforces degeneracies. Additional microscopic symmetries matter only when they modify these constraints or introduce new degeneracies. Since the symmetries distinguishing the AZ classes above do neither, they define no new MBL classes; rather, they reduce to the classes already identified 
 (see the End Matter).

 In the absence of symmetry constraints (class A), the emergent LIOMs
 form a complete set of mutually commuting quasi-local conserved
 operators without additional symmetry restrictions.
 No symmetry-enforced degeneracies arise, and resonant
 processes are parametrically suppressed at strong disorder, making this
 class the canonical setting for stable MBL.
 A representative realization is the disordered XXZ spin chain~\cite{classA1,classA2,classA3},
 \begin{equation}\notag 
 H_{A}=\sum_{i} \left[ J_z S_i^z S_{i+1}^z
 + J_\perp (S_i^x S_{i+1}^x + S_i^y S_{i+1}^y) 
 + h_i S_i^z\right],
 \end{equation}
 where $S_i^\alpha$ ($\alpha = x,y,z$) are spin-$1/2$ operators, $J_z$ and $J_\perp$ are the longitudinal and transverse exchange
 couplings, and $h_i\in[-W,W]$ are random fields.
 For $W\gg J_z,J_\perp$, perturbative constructions and numerical studies
 establish a robust MBL phase.
 
 Imposing time-reversal symmetry with $T^2=+1$ places the system in
 class AI. Time reversal acts by complex conjugation and remains compatible with
 the LIOM algebra, without enforcing local multiplets or
 symmetry-enforced degeneracies.
 Accordingly, the stability of MBL closely parallels that of class A. An illustrative example is the disordered spin chain~\cite{classAI1}
 \begin{equation}\notag 
 H=\sum_i\bigl(
 J_i\sigma_i^x\sigma_{i+1}^x
 +h_i\sigma_i^z
 +J'_i\sigma_i^z\sigma_{i+1}^z
 \bigr)
 \end{equation}
 with real random couplings $J_i$, $J'_i$, and $h_i$. When $\sigma_i^\alpha$ describe effective two-level degrees of freedom,
 time reversal acts as complex conjugation,
 $T=K$,
 placing the model in class AI. When $h_i=0$, an additional unitary $\mathbb{Z}_2$ symmetry
 $P=\prod_j\sigma_j^x$ emerges, yielding the AI$+\mathbb{Z}_2$ class and
 allowing SPT MBL phases in one dimension.  
 If instead $\sigma_i^\alpha$ represent physical spin-$1/2$ moments, time
 reversal is implemented as
 $T=\prod_j (i\sigma_j^y)K$, for which $T^2=-1$. The model therefore belongs to class AII, where the LIOM description
 remains valid but symmetry-enforced Kramers degeneracies render the MBL
 phase fragile.
 
 In symmetry class AIII, chiral symmetry is represented by a unitary
 operator $S$ satisfying $S H S^{-1} = -H$, which pairs many-body eigenstates at opposite energies.
 Unlike time-reversal symmetry with $T^2=-1$, this spectral pairing does
 not generally enforce exact degeneracies away from zero
 energy.
 Consequently, chiral symmetry remains compatible with stable MBL and may
 protect topological edge modes throughout the spectrum. 
 A representative example is the disordered interacting Dirac
 chain~\cite{SPT2},
 $H_{\mathrm{AIII}} =
 -\sum_{j} \left[J_j \left(
 c_j^\dagger c_{j+1}^\dagger + c_j^\dagger c_{j+1} + \mathrm{H.c.}
 \right)+h_j \left( 1 - 2c_j^\dagger c_j \right)\right]$,
 where $c_j$ annihilates a spinless fermion and $J_j,h_j$ are quenched
 random couplings. With weak symmetry-preserving interactions, the system admits a
 complete LIOM description while preserving chiral symmetry, thereby
 realizing a stable AIII-class MBL phase. 
 Within our framework, the interacting disordered SSH model likewise belongs to class AIII (see the End Matter).
 
 Particle-hole symmetric classes D ($C^2=+1$) and C ($C^2=-1$) describe
 superconducting systems formulated in terms of Bogoliubov--de Gennes
 (BdG) Hamiltonians.
 Class D MBL has been extensively studied in disordered interacting
 Kitaev chains and related Majorana systems~\cite{SPT2,Dclass},
 whereas explicit studies of class C MBL remain comparatively limited.
 Nevertheless, from the viewpoint of the present classification,
 particle-hole symmetry alone neither obstructs the construction of a
 complete LIOM algebra nor enforces 
 degeneracies. Stable MBL is therefore expected to be compatible with
 both classes D and C.
 A minimal interacting realization of class C is given by,
 \begin{align}
 H_{\text{C}} &=
 \sum_{i,\sigma}
 \left( t_i c_{i\sigma}^\dagger c_{i+1\sigma}
 + \text{h.c.} \right)
 + \sum_i
 \left( \Delta_i c_{i\uparrow}^\dagger c_{i\downarrow}^\dagger
 + \text{h.c.} \right) \nonumber \\
 &+ \sum_{i,\sigma} \mu_i c_{i\sigma}^\dagger c_{i\sigma}+\sum_i U_i
 \left( n_{i\uparrow}-\tfrac{1}{2} \right)
 \left( n_{i\downarrow}-\tfrac{1}{2} \right), \nonumber
 \end{align}
 where $c_{i\sigma}$ annihilates a fermion with spin
 $\sigma=\uparrow,\downarrow$ on site $i$, while $t_i,\Delta_i,\mu_i,U_i$ denote random hopping, pairing, chemical
 potential, and interaction strengths, respectively.
 This model possesses particle-hole symmetry with $C^2=-1$ and therefore belongs to class C.
 In the strongly disordered regime, this system is 
 expected to admit a
 complete LIOM description and therefore realize a stable
 MBL phase.
 
 A canonical realization of class A+$U(1)$ is provided by the disordered
 interacting spinless fermion chain,
 \begin{equation}\notag   
 H_{AU} =
 \sum_i \left[
 J_i \left( c_i^\dagger c_{i+1} + \text{h.c.} \right)
 + V_i n_i n_{i+1}
 + h_i n_i
 \right],
 \end{equation}
 where
 $J_i$, $V_i$, and $h_i$
 denote the nearest-neighbor hopping amplitudes,
 interaction strengths,
 and random onsite potentials,
 respectively.
 In general, the couplings $J_i$, $V_i$, and $h_i$ may be complex, so that
 the model possesses neither time-reversal nor particle-hole symmetry.
 It therefore belongs to class A, supplemented by a global U(1)
 particle-number conservation. In the strongly disordered regime, the LIOMs may be chosen to carry
 definite U(1) quantum numbers while remaining mutually commuting and
 quasi-local~\cite{SM}. Since U(1) symmetry does not enforce 
 degeneracies, the system realizes a stable MBL phase. 

 The remaining symmetry classes in Table~\ref{tbl:MBL_full} include those with additional onsite Abelian symmetries, such as
 AI$+$U(1), A$+\mathbb{Z}_2$, A$+\mathbb{Z}_2\times\mathbb{Z}_2$, and
 AI$+\mathbb{Z}_2\rtimes T$, as well as systems with continuous
 non-Abelian symmetries. Representative lattice models are given in the End Matter. 
 As established above, onsite Abelian symmetries are compatible with a
 complete LIOM algebra. U(1) charge conservation, for instance, requires
 LIOMs to carry definite charge quantum numbers, imposing selection
 rules without introducing degeneracies.
 Discrete Abelian symmetries like $\mathbb{Z}_2$ can act either
 linearly or projectively on the LIOM algebra. Projective actions give rise to distinct SPT MBL phases, while 
 linear ones lead to conventional stable MBL. A canonical example
 is the disordered cluster chain with
 $\mathbb{Z}_2\times\mathbb{Z}_2$ symmetry~\cite{SPT4}, whose bulk remains
 MBL while symmetry-protected edge modes persist across the many-body spectrum. By contrast, continuous non-Abelian symmetries, such as SU(2) or
 SU($N$), are incompatible with a complete commuting LIOM algebra.
 They enforce local multiplet structures that generate symmetry-enforced degeneracies and strongly enhance resonant processes, thereby precluding stable MBL.

 
 {\em Conclusion and Discussion.--}
 We have established a unified operator-algebraic framework for understanding symmetry constraints in MBL phases. Within this
 framework, the stability of MBL under a given symmetry is determined
 by two complementary criteria: whether the symmetry is compatible
 with a complete LIOM algebra and, if so, whether it enforces degeneracies. These criteria naturally classify MBL phases into stable, fragile, SPT, and unstable classes, while placing AZ symmetries together with onsite Abelian and continuous non-Abelian symmetries within a unified framework.
 
 Our classification is formulated entirely at the level of the emergent LIOM algebra. The stability criteria identified above therefore depend on the symmetry action on the emergent operator algebra rather than on the microscopic mechanism responsible for localization. 
 Besides quenched disorder, MBL may also arise from Stark potentials~\cite{StarkMBL1,StarkMBL2,StarkMBL3,StarkMBL4}. More generally, MBL can also occur in periodically driven Floquet systems~\cite{FloquetMBL,FloquetMBL1}. Whenever the corresponding effective description admits a complete set of LIOMs~\cite{StarkMBL4,FloquetMBL1} (see the Supplemental Material~\cite{SM}), the same symmetry-based criteria can be applied to the emergent LIOM algebra. Thus, the present classification applies to these systems.      
 
 Finally, we emphasize that the present symmetry classification is
 presented for one-dimensional systems. 
 The existence and ultimate stability of a complete LIOM description in higher dimensions may depend on the spatial
 dimensionality and the microscopic localization mechanism~\cite{2DMBL1,2DMBL2}.
 Should stable MBL be established beyond one dimension, the present framework can be naturally extended by incorporating spatial dimensionality as an additional classification index, analogous to its role in the AZ classification of topological phases. The present operator-algebraic framework provides a natural starting point for future studies of symmetry, MBL, and nonequilibrium many-body dynamics.



 \begin{acknowledgments}
 This work is supported by National Key R\&D Program of China under Grant No.2022YFA1405800.
 \end{acknowledgments}

 \section{End Matter}
 
 {\em Reduction of AZ Classes in the MBL Regime.---} 
 In the AZ classification of non-interacting fermionic
 systems, the ten symmetry classes are distinguished by different
 combinations of time-reversal ($T$), particle-hole ($C$), and chiral ($S$)
 symmetries. Classes such as BDI, DIII, CI, and CII are therefore genuinely distinct at
 the single-particle level and play an essential role in determining the
 structure of Anderson localization, nonlinear sigma models, and
 topological band theory. In the interacting MBL setting considered here, however, the organizing principle is fundamentally different.
 MBL is characterized by the existence of an extensive set of LIOMs, and the classification of MBL phases is governed by the operator-level constraints that global symmetries impose
 on the LIOM algebra. 
 When formulated at the level of LIOMs, several AZ symmetry classes that
 are distinct microscopically no longer generate independent localization
 constraints.
 In particular, the additional symmetry relations that distinguish the
 BDI, DIII, CI, and CII classes do not introduce new symmetry-enforced
 local degeneracies, Kramers multiplets, or independent operator
 constraints beyond those already represented by classes A, AI, AII,
 AIII, D, and C. Consequently, although these AZ classes remain distinct microscopic
 symmetry classes, they do not define independent MBL classes within our framework.
 
 More concretely, class BDI is characterized by $T^2=+1$, $C^2=+1$, and the resulting chiral symmetry
 $S=TC$. Since neither $T^2=+1$ nor $C^2=+1$ enforces Kramers degeneracies or
 symmetry-protected local multiplets, these symmetries constrain the
 LIOM algebra only weakly.
 Generically, if particle-hole symmetry does not induce a genuine
 Bogoliubov-de Gennes (BdG) structure, the resulting localization criterion reduces to that of class AI, whereas in the presence of a BdG structure it reduces to that of class D.
 Accordingly, generic BDI systems reduce to AI or D in the LIOM-based
 MBL classification, depending on their microscopic realization.
 A nongeneric but physically important exception occurs when chiral
 symmetry provides the only nontrivial constraint on the LIOM algebra,
 while time-reversal and particle-hole symmetries do not generate
 additional local constraints.
 In this situation, the localization criterion is completely determined
 by the chiral pairing structure, and the system is effectively described
 by class AIII within the present framework.
 A paradigmatic example is the Su-Schrieffer-Heeger (SSH) model.
 Although the noninteracting SSH model belongs to class BDI in the AZ
 classification, its LIOM algebra in the interacting MBL regime is
 governed solely by chiral symmetry, so that BDI and AIII exhibit the
 same localization criterion.
 
 Class DIII is characterized by $T^2=-1$ and $C^2=+1$. Within the LIOM framework, the essential localization constraint arises
 from the Kramers degeneracy enforced by $T^2=-1$.
 By contrast, particle-hole symmetry with $C^2=+1$ does not generate
 additional local degeneracies or independent operator constraints.
 Therefore, the localization criterion of class DIII reduces to that of
 class AII. Similarly, class CI is characterized by
 $T^2=+1$ and $C^2=-1$.
 The time-reversal symmetry with $T^2=+1$ does not enforce Kramers
 degeneracies or local symmetry multiplets.
 The remaining particle-hole symmetry determines the BdG structure but
 does not introduce an additional localization constraint beyond that
 already represented by class C.
 Accordingly, the localization criterion of class CI reduces to that of
 class C. Finally, class CII combines
 $T^2=-1$ and $C^2=-1$.
 Within the present LIOM framework,
 the particle-hole symmetry determines the
 Bogoliubov-de Gennes (BdG) structure,
 whereas the time-reversal symmetry with
 $T^2=-1$ enforces local Kramers degeneracies.
 According to the classification principle established in the main text,
 when multiple symmetry constraints coexist,
 the effective MBL class is determined by the constraint that provides
 the strongest operator-level obstruction to stable LIOMs.
 Since the Kramers degeneracy already leads to the fragile-MBL criterion
 represented by class AII,
 the additional BdG structure does not modify the resulting localization
 stability.
 Accordingly, class CII reduces to class AII within the present
 LIOM-based MBL classification.

 In summary, BDI, DIII, CI, and CII remain distinct symmetry classes of
 microscopic Hamiltonians, but they do not define independent classes of
 MBL phases once the problem is formulated in terms of quasi-local LIOMs.
 The resulting reduction in the LIOM-based MBL classification is
 summarized below:
 \begin{center}
 \begin{tabular}{cc}
 	\hline
 	AZ symmetry class &
 	\qquad MBL classification \\
 	\hline
 	BDI  & \qquad AI or D (generically)\\
 	& \qquad AIII (special case)\\
 	DIII & \qquad AII\\
 	CI   & \qquad C\\
 	CII  & \qquad AII\\
 	\hline
 \end{tabular}
 \end{center}
 This table summarizes the reduction of localization criteria within the
 LIOM-based MBL classification rather than a reduction of the
 AZ symmetry classes themselves. The reduction follows from the classification principle: when multiple symmetry constraints coexist, the effective MBL class is determined by the symmetry constraint that provides the strongest operator-level obstruction to stable LIOMs.
 Different AZ symmetry classes may therefore correspond to the same
 localization criterion whenever the additional symmetry relations do not
 introduce new operator-level constraints on the LIOM algebra.
 
 {\em Representative Models for Symmetry Classes Not Illustrated in the Main Text.--} In the main text, several symmetry classes appearing in
 Table~\ref{tbl:MBL_full} have already been illustrated by explicit
 microscopic lattice models.
 In this section, we complement that discussion by providing minimal
 representative models for the remaining classes, in particular those
 involving additional onsite Abelian symmetries beyond the standard
 Altland--Zirnbauer (AZ) structure.
 These examples are not intended to be exhaustive, but rather to
 demonstrate that the corresponding symmetry constraints can be realized
 in concrete interacting disordered systems and analyzed consistently
 within the LIOM framework.

 A representative model for class AI$+U(1)$ is a disordered interacting
 fermionic chain with an orbital (rather than physical-spin) degree of
 freedom carrying a time-reversal symmetry satisfying $T^2=+1$. A minimal example is 
 \begin{equation}\notag 
 H =
 \sum_{i,\sigma}
 t_i \left( c_{i\sigma}^\dagger c_{i+1\sigma} + \text{h.c.} \right)
 + \sum_i U_i n_{i1} n_{i2}
 + \sum_{i,\sigma} \mu_i n_{i\sigma}.
 \end{equation}
 where $\sigma=1,2$ labels two orbital (or pseudospin) flavors rather
 than the physical electron spin.
 The time-reversal operator acts trivially on the orbital index, so that
 $T^2=+1$.
 Time-reversal symmetry therefore does not enforce Kramers degeneracies,
 while the conserved U(1) charge simply requires the LIOMs to remain
 charge resolved.
 Consequently, this class supports stable MBL phases, analogous to
 class A$+U(1)$.
 
 Discrete onsite symmetries can also be incorporated within the LIOM
 framework.
 A simple example of class A$+\mathbb{Z}_2$ is provided by a disordered
 interacting Ising chain,
 \begin{equation}\notag 
 H =
 \sum_i
 J_i \sigma_i^z \sigma_{i+1}^z
 + h_i \sigma_i^x
 + J_i' \sigma_i^x \sigma_{i+1}^x,
 \end{equation}
 which is invariant under the global $\mathbb{Z}_2$ transformation
 $P=\prod_i \sigma_i^x$.
 Depending on how the symmetry acts on the LIOMs, such systems may
 realize either symmetry-preserving trivial MBL phases or
 SPT MBL phases.
 
 As a representative realization of the class
 A$+\mathbb{Z}_2 \times \mathbb{Z}_2$ entry in
 Table~\ref{tbl:MBL_full}, a paradigmatic example of a
 SPT-MBL phase with discrete onsite
 symmetry is provided by the disordered cluster model~\cite{SPT4},
 \begin{equation}\notag 
 H =
 \sum_i J_i \sigma_{i-1}^z \sigma_i^x \sigma_{i+1}^z
 + \sum_i h_i \sigma_i^x,
 \end{equation}
 which is invariant under independent $\mathbb{Z}_2$ symmetries acting
 on the even and odd sublattices.
 In this model, the bulk degrees of freedom are fully localized, while
 the emergent LIOMs transform projectively under
 $\mathbb{Z}_2 \times \mathbb{Z}_2$.
 As a result, symmetry-protected edge modes persist at all energies in
 the spectrum, providing a canonical example of an SPT-MBL phase. 
 
 A representative realization of class AI$+\mathbb{Z}_2 \rtimes T$
 can be constructed from a disordered spin-1 chain,
 \begin{equation}\notag 
 H =
 \sum_i J_i \mathbf{S}_i \cdot \mathbf{S}_{i+1}
 + \sum_i D_i (S_i^z)^2 ,
 \end{equation}
 where $J_i$ and $D_i$ are independent random couplings.
 This Hamiltonian is inspired by Ref.~\cite{EndMBL}, although
 the disorder realization and symmetry implementation are different.
 The model preserves time-reversal symmetry satisfying
 $T^2=+1$.
 In addition, it possesses the onsite
 $\mathbb Z_2$
 symmetry generated by $P=\prod_i e^{i\pi S_i^x}$. Together,
 $P$ and $T$
 generate the semidirect product
 $\mathbb Z_2\rtimes T$.
 At the level of quasi-local integrals of motion,
 this symmetry permits projective edge representations while remaining
 compatible with stable MBL in one dimension.

\global\long\def\id{\mathbbm{1}}
\global\long\def\ui{\mathbbm{i}}
\global\long\def\ud{\mathrm{d}}
\setcounter{equation}{0} \setcounter{figure}{0}
\setcounter{table}{0} 
\renewcommand{\theparagraph}{\bf}
\renewcommand{\thefigure}{S\arabic{figure}}
\renewcommand{\theequation}{S\arabic{equation}}

\onecolumngrid
\flushbottom
\newpage
\section*{\large Supplementary Material:\\ Unified Symmetry Classification of Many-Body Localized Phases}

In the Supplementary Material, we first review the Schrieffer-Wolff (SW) construction of local integrals of motion (LIOMs) and illustrate their emergence through representative examples. We then provide the detailed derivation of Eq.~(4) in the main text and present a possible physical interpretation of fragile many-body localization (MBL) in class AII. Finally, we discuss the applicability of the present classification beyond conventional disorder-induced MBL by considering both Stark MBL systems and Floquet MBL systems.

\section{I. Construction of LIOMs}
In this section, we review the SW construction of
LIOMs in strongly disordered interacting systems. We use representative
spin and fermion models to illustrate how LIOMs emerge perturbatively,
how microscopic symmetries are inherited by the emergent LIOM algebra,
and how different realizations of U(1) symmetry lead to distinct
operator-algebraic classifications.

\subsection{A. SW Construction of LIOMs}
We consider a generic strongly disordered lattice Hamiltonian 
\[
H = H_0 + \lambda V ,
\]
where $H_0$ is diagonal in a local product basis and $\lambda\ll1$
controls the strength of the off-diagonal perturbation.
For concreteness, one may take $H_0 = \sum_i h_i \sigma_i^z$, 
with strong random fields $h_i \in [-W, W]$ satisfying
$W\gg \lambda\|V\|$. The perturbation is decomposed into diagonal and off-diagonal parts with
respect to the eigenbasis of $H_0$,
\[
V = V_{\mathrm{d}} + V_{\mathrm{od}},
\]
where $[H_0, V_{\mathrm{d}}]=0$ and $V_{\mathrm{od}}$ 
contains all matrix elements connecting different eigenstates of
$H_0$. The SW transformation seeks a quasi-local unitary
transformation $U = e^{S}$, with $S^\dagger = -S$, such that the transformed Hamiltonian
$\tilde H = e^{S} H e^{-S}$
is diagonal order by order in perturbation theory.

The generator $S$ is expanded perturbatively as
\[
S = \sum_{n=1}^{\infty} \lambda^n S^{(n)} .
\]
Using the Baker--Campbell--Hausdorff expansion,
\[
\tilde H
= H + [S,H]
+ \frac{1}{2}[S,[S,H]]
+ \frac{1}{3!}[S,[S,[S,H]]] + \cdots ,
\]
we impose that the off-diagonal part of $\tilde H$ vanishes order by order in $\lambda$. At first order, the generator $S^{(1)}$ is fixed by the condition
\[
[S^{(1)}, H_0] = - V_{\mathrm{od}} ,
\]
which, in the eigenbasis of $H_0$, $H_0|\alpha\rangle = E_\alpha|\alpha\rangle$,
yields
\[
\langle \alpha | S^{(1)} | \beta \rangle
= \frac{-\langle \alpha | V_{\mathrm{od}} | \beta \rangle}{E_\alpha - E_\beta},
\qquad E_\alpha \neq E_\beta .
\]
Introducing the Liouvillian superoperator
$\mathcal L_{H_0}(X)=[H_0,X]$,
the formal solution may be written as
\[
S^{(1)}
=
-\mathcal L_{H_0}^{-1}(V_{\mathrm{od}}),
\]
where
$\mathcal L_{H_0}^{-1}$
acts only on off-diagonal operators. This inverse Liouvillian $\mathcal{L}_{H_0}^{-1}$ is well defined provided no exact degeneracy is present within the subspace connected by
$V_{\mathrm{od}}$.
At second order, $S^{(2)}$ is chosen to cancel the off-diagonal contributions generated by commutators involving $S^{(1)}$,
\[
[S^{(2)}, H_0]
= -\mathcal{P}_{\mathrm{od}}
\left(
\frac{1}{2}[S^{(1)},V]
+ \frac{1}{2}[S^{(1)},[S^{(1)},H_0]]
\right),
\]
where $\mathcal{P}_{\mathrm{od}}$ projects onto the off-diagonal subspace. Higher orders follow recursively.

\subsection{B. Example 1: Disordered XXZ Spin Chain (Class A)}
To make the construction explicit, we first apply this procedure to the random XXZ spin chain,
\[
H_{\mathrm{XXZ}}
= \sum_{i=1}^{L} h_i S_i^z
+ \sum_{i=1}^{L-1}
\left[
J_z S_i^z S_{i+1}^z
+ \frac{J_\perp}{2}
\left(
S_i^+ S_{i+1}^- + S_i^- S_{i+1}^+
\right)
\right],
\]
where $S_i^\pm = S_i^x \pm i S_i^y$, $h_i$ are independent random variables uniformly distributed in $[-W, W]$, and $J_z, J_\perp \sim J$ with $J/W \ll 1$. We identify
\[
H_0 = \sum_i h_i S_i^z,
\qquad
V = \sum_i
\left[
J_z S_i^z S_{i+1}^z
+ \frac{J_\perp}{2}
\left(
S_i^+ S_{i+1}^- + S_i^- S_{i+1}^+
\right)
\right],
\]
and treat $\lambda \sim J/W$ as the small parameter.

The off-diagonal part of $V$ is
\[
V_{\mathrm{od}}
= \sum_i \frac{J_\perp}{2}
\left(
S_i^+ S_{i+1}^- + S_i^- S_{i+1}^+
\right).
\]
Solving $[S^{(1)},H_0] = -V_{\mathrm{od}}$ yields
\[
S^{(1)}
= \sum_i
\frac{J_\perp}{2(h_i - h_{i+1})}
\left(
S_i^+ S_{i+1}^- - S_i^- S_{i+1}^+
\right).
\]
The energy denominators $(h_i-h_{i+1}) \sim W$ ensure that $S^{(1)}$ is perturbatively small.

Keeping terms up to second order, the transformed Hamiltonian becomes
\[
\tilde H
= \sum_i h_i S_i^z
+ \sum_i J_z S_i^z S_{i+1}^z
+ \frac{1}{2}
\sum_i
\frac{J_\perp^2}{2(h_i-h_{i+1})}
\left(S_i^z - S_{i+1}^z\right)
+ \mathcal{O}(J^3/W^2),
\]
where the second-order contribution $\frac{1}{2}[S^{(1)},V_{\mathrm{od}}]$ generates additional diagonal corrections.

The emergent LIOMs are obtained by dressing the bare local operators,
\[
\tau_i^z = e^{S} S_i^z e^{-S}.
\]
To first order,
\[
\tau_i^z
\simeq S_i^z + [S^{(1)},S_i^z]
= S_i^z
+ \frac{J_\perp}{2(h_i-h_{i+1})}
\left(
S_i^+ S_{i+1}^- - S_i^- S_{i+1}^+
\right)
+ \frac{J_\perp}{2(h_i-h_{i-1})}
\left(
S_i^+ S_{i-1}^- - S_i^- S_{i-1}^+
\right).
\]
This expression illustrates how virtual spin-flip processes dress an
initially local operator, providing the leading-order structure of the
emergent LIOM.

For strong disorder, typical matrix elements satisfy
\[
\frac{|V_{\alpha\beta}|}{|E_\alpha - E_\beta|}
\sim \frac{J}{W} < 1 ,
\]
ensuring convergence of the SW expansion. Higher-order corrections involve increasingly long virtual processes, whose amplitudes are controlled by successive energy denominators of
order $W$. Consequently, the dressing of the LIOMs remains exponentially localized
when the perturbative expansion converges.

\subsection{C. Example 2: Disordered Spinless Fermion Chain (Class A+U(1))}
We next consider a disordered interacting spinless fermion chain, which
provides a representative realization of class A+U(1), 
\[
H_{\mathrm{ferm}}
= \sum_i \left[
t_i (c_i^\dagger c_{i+1} + \mathrm{h.c.})
+ V_i n_i n_{i+1}
+ \mu_i n_i
\right],
\]
where $n_i = c_i^\dagger c_i$, and $t_i$, $V_i$, $\mu_i$ are random with typical magnitude $|t_i|, |V_i| \ll |\mu_i| \sim W$. This Hamiltonian conserves total particle number, which generates the global U(1) symmetry.

We take $H_0 = \sum_i \mu_i n_i$ and
\[
V = \underbrace{\sum_i t_i (c_i^\dagger c_{i+1} + \mathrm{h.c.})}_{V_{\mathrm{od}}}
+ \underbrace{\sum_i V_i n_i n_{i+1}}_{V_{\mathrm{d}}}.
\]

The first-order generator must satisfy $[S^{(1)}, H_0] = -V_{\mathrm{od}}$. For a nearest-neighbour hopping process, the corresponding energy
difference is $E_\alpha-E_\beta
=
\mu_{i+1}-\mu_i$. Thus
\[
S^{(1)} = \sum_i \frac{t_i}{\mu_{i+1} - \mu_i} (c_i^\dagger c_{i+1} - c_{i+1}^\dagger c_i).
\]

The transformed Hamiltonian is
\[
\tilde H = \sum_i \mu_i n_i + \sum_i V_i n_i n_{i+1}
+ \frac{1}{2} \sum_i \frac{t_i^2}{\mu_{i+1}-\mu_i} (n_i - n_{i+1})
+ \mathcal{O}(t^3/W^2).
\]

The corresponding LIOMs are naturally chosen as dressed local number
operators, $\tau_i = e^{S} n_i e^{-S}$. To first order:
\[
\tau_i \simeq n_i + [S^{(1)}, n_i]
= n_i + \frac{t_{i-1}}{\mu_i - \mu_{i-1}} (c_{i-1}^\dagger c_i + c_i^\dagger c_{i-1})
- \frac{t_i}{\mu_{i+1} - \mu_i} (c_i^\dagger c_{i+1} + c_{i+1}^\dagger c_i).
\]

\subsection{D. Comparison of Symmetry Realization in the LIOM Algebra}
The two examples illustrate that the same microscopic U(1) symmetry
need not play the same role in the construction of the emergent LIOM
algebra.
For the XXZ spin chain, the perturbative construction naturally starts
from the local operators $S_i^z$, which provide a convenient set of bare
operators in the strong-disorder limit.
Although the microscopic Hamiltonian conserves the total spin
$S_{\mathrm{tot}}^z=\sum_iS_i^z$, this conservation law does not impose
an independent structural constraint on the LIOM algebra.
The SW construction can be carried out entirely without explicit
reference to the U(1) generator.
Only after the LIOMs have been constructed does one recover the
conserved quantity through
$S_{\mathrm{tot}}^z=\sum_i\tau_i^z$,
up to quasi-local dressing.
From the perspective of the present classification, the U(1) symmetry is
therefore absorbed into the LIOM construction rather than
defining a distinct operator-algebraic structure.

By contrast, for the interacting spinless fermion chain, particle-number
conservation constrains the LIOM construction from the outset.
The admissible local operators must respect the U(1) superselection
structure, and the dressed LIOMs are required to remain charge-neutral
operators that preserve particle number.
Consequently, the emergent LIOM algebra is naturally organized according
to definite U(1)-charge sectors.
Here the U(1) symmetry is not simply recovered after the LIOMs are
constructed, but constitutes an intrinsic structural constraint on the
operator algebra itself.
It is this distinction that motivates the separate class A$+$U(1) in the
present framework.

The different roles played by U(1) symmetry become particularly evident
when the symmetry is weakly broken.
For the XXZ chain, generic symmetry-breaking perturbations modify the
microscopic Hamiltonian but do not qualitatively alter the scalar
structure of the emergent LIOM algebra.
The resulting MBL phase therefore remains continuously connected to that
of class A.
By contrast, breaking particle-number conservation in the spinless
fermion system generally requires enlarging the operator algebra and
reconstructing the LIOMs in terms of Bogoliubov quasiparticles.
The effective symmetry class consequently changes, typically to a Bogoliubov-de Gennes (BdG)
class such as D or C depending on the remaining symmetries.
This illustrates that, within the present classification, the crucial
issue is not the presence of a microscopic U(1) symmetry itself, but
whether that symmetry constitutes an independent structural constraint
on the emergent LIOM algebra.

\subsection{E. General Symmetry Constraints on the LIOM Algebra}

We now show how global symmetries are inherited by the emergent LIOM
algebra within the perturbative SW construction.
Let $G$ be a global symmetry group represented by unitary or
anti-unitary operators $g$ such that both the unperturbed Hamiltonian
$H_0$ and the perturbation $V$ are invariant under the symmetry,
\[
gH_0g^{-1}=H_0,
\qquad
gVg^{-1}=V.
\]
Since $V_{\mathrm{od}}$ is obtained by projecting $V$ onto the
off-diagonal subspace of $H_0$, it follows that
$gV_{\mathrm{od}}g^{-1}=V_{\mathrm{od}}$.
The recursive equations defining the SW generators
are therefore covariant under the symmetry transformations.

At first order, the SW generator satisfies 
$[S^{(1)},H_0]=-V_{\mathrm{od}}$.
Applying the symmetry transformation gives
\[
[gS^{(1)}g^{-1},\,H_0]
=
-V_{\mathrm{od}},
\]
which is identical to the original defining equation.
Fixing the gauge by requiring the diagonal part of $S^{(1)}$ to vanish
removes the freedom to add operators commuting with $H_0$, so the
solution is unique and therefore
\[
gS^{(1)}g^{-1}=S^{(1)}.
\]
The same argument applies recursively at every perturbative order:
the transformed generator satisfies the same defining equation and the
same gauge condition as the original one, and therefore
$gS^{(n)}g^{-1}=S^{(n)}$ 
for all orders for which the SW construction is well defined.

The covariance of the LIOMs follows directly from their definition,
$\tau_i^z=e^SO_i e^{-S}$,
where $S=\sum_{n\geq1}S^{(n)}$ and $O_i$ denotes the corresponding microscopic local operator.
Applying a symmetry transformation yields
\[
g\tau_i^zg^{-1}
=
ge^Sg^{-1}\,
gO_i g^{-1}\,
ge^{-S}g^{-1}.
\]
Since $gSg^{-1}=\pm S$,
one has
\[
ge^Sg^{-1}=e^{\pm S},
\]
showing that the transformed LIOM is obtained by dressing the
transformed microscopic operator in exactly the same manner.
Hence the emergent LIOM algebra inherits the symmetry transformation
properties of the underlying microscopic degrees of freedom.

The perturbative SW construction therefore provides a concrete
illustration of the general principle adopted in the main text:
whenever a complete LIOM algebra exists, the microscopic
symmetry acts consistently on the emergent operator algebra.
This operator-algebraic viewpoint naturally distinguishes symmetries
that are compatible with a complete LIOM algebra from those that are
not.
For continuous non-Abelian symmetries such as SU(2), the symmetry
requires local operators to transform as nontrivial multiplets.
Such multiplet structures are incompatible with a complete commuting
quasi-local LIOM algebra, in agreement with the no-go theorems for
stable MBL in the presence of continuous non-Abelian symmetries.

\section{II. Derivation of Eq.~(4)}

For completeness, we derive the transformation law used in Eq.~(4) of the main text.
Suppose, for contradiction, that a complete LIOM algebra exists while preserving SU(2) symmetry. The corresponding LIOM operators are then required to transform as a
vector representation of SU(2),
\begin{equation}
	U(R)\tau_i^\alpha U^\dagger(R)
	=
	R_{\alpha\beta}\tau_i^\beta,
	\label{eq:vector_rep}
\end{equation}
where $U(R)$ denotes the unitary representation of an SU(2) rotation acting on the Hilbert space and  $R_{\alpha\beta}$ is the corresponding SO(3) rotation matrix acting on the vector operator
$(\tau_i^x,\tau_i^y,\tau_i^z)^T$. 

For a rotation by an angle $\theta$ about the $y$ axis, 
\begin{equation}\notag 
	R_y(\theta)
	=
	\begin{pmatrix}
		\cos\theta & 0 & -\sin\theta\\
		0 & 1 & 0\\
		\sin\theta & 0 & \cos\theta
	\end{pmatrix}.
\end{equation}
Equation~(\ref{eq:vector_rep}) therefore implies 
\begin{equation}\notag
	U(R_y)
	\begin{pmatrix}
		\tau_i^x\\
		\tau_i^y\\
		\tau_i^z
	\end{pmatrix}
	U^\dagger(R_y)
	=
	R_y(\theta)
	\begin{pmatrix}
		\tau_i^x\\
		\tau_i^y\\
		\tau_i^z
	\end{pmatrix},
\end{equation}
from which the third component immediately gives
\begin{equation}\notag 
	e^{i\theta S_y}
	\tau_i^z
	e^{-i\theta S_y}
	=
	\cos\theta\,\tau_i^z
	-
	\sin\theta\,\tau_i^x,
\end{equation}
which is Eq.~(4) of the main text.

The important point is that this transformation law is not an
additional assumption.
Rather, it is a necessary consequence of the hypothesis that a complete
LIOM algebra exists and remains compatible with SU(2) symmetry.
Since SU(2) rotations necessarily mix different components of each local
operator multiplet, no individual scalar LIOM can remain invariant under
the full symmetry.
This illustrates why continuous non-Abelian symmetries are incompatible
with a complete commuting LIOM algebra and therefore obstruct stable
MBL, consistent with the no-go theorems discussed in the main text.

\section{III. Physical Interpretation of Fragile MBL in Class AII}

In the main text, we showed that symmetry class AII differs
fundamentally from systems with continuous non-Abelian symmetries.
Unlike continuous non-Abelian symmetries, time-reversal symmetry with
$\mathcal T^2=-1$ remains compatible with a complete LIOM
algebra.
Class AII is nevertheless distinguished from conventional stable MBL by
the presence of symmetry-enforced Kramers degeneracies.

To the best of our knowledge, although previous works have argued that
Kramers degeneracy may ultimately destabilize MBL, the microscopic
origin of this instability has not been discussed in detail.
Here we present a possible physical interpretation.
Within the SW construction, the quasi-local unitary transformation is
generated perturbatively order by order.
At the $n$th order, the effective coupling between two many-body
configurations $|\alpha\rangle$ and $|\beta\rangle$ arises through a
sequence of virtual intermediate configurations
\[
|\alpha\rangle
\rightarrow
|\gamma_1\rangle
\rightarrow
\cdots
\rightarrow
|\gamma_n\rangle
\rightarrow
|\beta\rangle,
\]
where $|\alpha\rangle$, $|\beta\rangle$, and
$\{|\gamma_k\rangle\}$ denote many-body basis states of the unperturbed
Hamiltonian.
Schematically, the corresponding $n$th-order transition amplitude may be
written as
\begin{equation}\notag
	A_{\alpha\beta}^{(n)}
	\sim
	\prod_{k=1}^{n}
	\frac{
		\langle\gamma_k|V|\gamma_{k-1}\rangle
	}{
		E_{\gamma_k}-E_{\gamma_{k-1}}
	},
	\label{eq:SW}
\end{equation}
where $V$ denotes the perturbation generating the virtual transitions and
$E_{\gamma_k}$ is the unperturbed energy of the intermediate state
$|\gamma_k\rangle$.
In conventional MBL, the typical energy denominators are of order $W$,
so higher-order virtual processes are strongly suppressed and the SW
expansion remains quasi-local.

In class AII, however, every many-body configuration has a
Kramers partner with exactly the same energy.
Unlike the local multiplet structure generated by continuous
non-Abelian symmetries, these degeneracies do not obstruct a complete LIOM algebra, nor do they produce an exponentially large
local degeneracy $\mathcal D=\prod_{i=1}^{N}d_i$,
where $d_i$ is the dimension of the local multiplet at the
$i$-th LIOM and $N$ is the number of LIOMs.
For non-Abelian symmetries, $d_i>1$ generically, leading to an
exponential degeneracy $\mathcal D$.
By contrast, Kramers degeneracy is always twofold. Although the number
of Kramers pairs grows with system size, the degeneracy
associated with each pair remains fixed and does not scale with system
size. Nevertheless, exact Kramers degeneracies increases the
number of virtual processes with anomalously small energy
denominators.
As the system size grows, the Hilbert-space dimension, and hence the
number of high-order virtual processes, increases exponentially.
Although each near-resonant process remains individually rare, the total
number of such processes grows rapidly, making it increasingly likely
that multiple resonances become interconnected to form extended
resonant clusters.  Such clusters facilitate avalanche-like thermalization and may ultimately destabilize localization on asymptotically large scales. This picture explains why class AII can show  MBL behavior in
finite-size systems and over experimentally accessible evolution times,
while becoming unstable in the thermodynamic and infinite-time
limits.

These considerations motivate our terminology.  
Unlike continuous non-Abelian symmetry classes, class AII remains
compatible with a complete LIOM algebra and therefore
retains the defining operator-algebraic structure of MBL.
However, the symmetry-enforced Kramers degeneracies distinguish it from
conventional stable MBL phases and may ultimately compromise its
asymptotic stability.
For this reason, we classify class AII as a fragile MBL phase.

\section{IV. Applicability of the Present MBL Classification}
The MBL classification developed in the main text is formulated entirely
at the level of the emergent LIOM algebra rather than at the level of
the microscopic mechanism responsible for localization.
Within this framework, the stability of MBL is determined by two
complementary criteria: whether the symmetry admits a complete LIOM
algebra and whether it enforces degeneracies.
These criteria depend only on the symmetry action on the emergent
operator algebra and therefore do not depend on whether the LIOM
description arises from quenched disorder, Stark localization, or
periodic driving.
Consequently, whenever a complete LIOM description exists,
the same symmetry classification applies irrespective of the
microscopic mechanism responsible for localization.
In this section, we illustrate this applicability beyond conventional
disorder-induced MBL by considering Stark and Floquet
MBL systems.

\subsection{A. Stark Localization within the Present MBL Classification}

Recent work has shown that Stark MBL systems can
admit an effective LIOM description in the strong-field
regime, see e.g.~\cite{DynamicalLbits}.
As a concrete example, consider the tilted XXZ spin chain
\begin{equation}\notag 
	H =
	\frac{J}{2}\sum_{j=1}^{L-1}
	\left(
	S_j^+S_{j+1}^-+
	S_j^-S_{j+1}^+
	\right)
	-\Delta\sum_{j=1}^{L-1} S_j^zS_{j+1}^z
	-W\sum_{j=1}^L jS_j^z ,
\end{equation}
where $J$ is the hopping amplitude, $\Delta$ is the 
interaction strength, and $W$ is the Stark-field gradient. In the
strong-field regime,
$W\gg J,\Delta$, the
SW transformation eliminates the hopping 
term order by order. To second order in $J/W$, the resulting effective
Hamiltonian takes the form~\cite{DynamicalLbits} 
\begin{equation}\notag 
	H_{\rm eff}=-W\sum_j jS_j^z
	-\Delta\sum_j S_j^zS_{j+1}^z
	+\frac{J^2}{4W}
	\left(S_L^z-S_1^z\right)
	+\mathcal{O}\left(\frac{J^3}{W^2}\right),
\end{equation}
which is diagonal in
the $S^z$ basis up to corrections that are perturbatively small in
$J/W$. The operators $S_j^z$ are therefore conserved by the effective
Hamiltonian, $[H_{\rm eff},S_j^z]=0$,
and provide the local building blocks of the emergent LIOM algebra.
If $U$ denotes the quasi-local unitary transformation relating the
effective and microscopic descriptions,
$H_{\rm eff}=U^\dagger H U$,
the corresponding conserved LIOMs of the microscopic Hamiltonian are
$\tau_j^z=U S_j^zU^\dagger$,
which satisfy $[H,\tau_j^z]=0$.
Thus, the strong-field expansion provides a direct construction of a
complete LIOM algebra.

It is useful to distinguish these conserved LIOMs from the
dynamical l-bits discussed in Ref.~\cite{DynamicalLbits}.
The latter are quasi-local raising and lowering operators, which can be
chosen as dressed versions of local spin-flip operators. They satisfy
relations of the form
$[H,\tau_j^\pm]=\pm\omega_j\tau_j^\pm$,
where the
corresponding frequencies $\omega_j$ include $jW$ and $jW\pm\Delta$~\cite{DynamicalLbits}. These dynamical l-bits
remain quasi-local and are exponentially long-lived in the strong-field
regime. Importantly, they are not to be confused with the conserved
$S_j^z$-type LIOMs: the former describe local dynamical excitations,
whereas the latter form the commuting algebra that labels the emergent
many-body eigenstates.

The tilted XXZ chain possesses a global U(1) symmetry corresponding to
the conservation of total magnetization,
$S_{\rm tot}^z=\sum_j S_j^z$.
Since this symmetry acts on the emergent LIOM algebra in the same way as
in the corresponding disorder-induced system, the Stark field does not
introduce a new symmetry class within the present framework. 

\subsection{B. Extension of the Present MBL Classification to Floquet MBL}

The MBL classification is formulated at the level of the emergent LIOM
algebra, and this viewpoint naturally extends to periodically driven
systems. Consider a periodically driven Hamiltonian
$H(t)=H(t+T)$,
with driving period $T$. The evolution over one period is described by
the Floquet operator
$F=\mathcal{T}
\exp\left(
-i\int_0^T H(t)\,dt
\right)$,
where $\mathcal{T}$ denotes time ordering. In a Floquet MBL phase, there
exists a complete set of LIOMs
$\{\tau_i^z\}$ satisfying \cite{FloquetMBL1s}
\begin{equation}\notag
	[F,\tau_i^z]=0,
	\qquad
	[\tau_i^z,\tau_j^z]=0,
	\qquad
	(\tau_i^z)^2=1,
\end{equation}
which can be used to label the Floquet many-body eigenstates. After choosing a consistent quasienergy branch, the Floquet operator may
be written as
$F=e^{-iH_{\rm eff}}$,
where the effective Floquet Hamiltonian $H_{\rm eff}$ is diagonal in the
LIOM basis and can be expressed in the quasi-local form~\cite{FloquetMBL1s}
\begin{equation}\notag
	H_{\rm eff}
	=
	\sum_i h_i \tau_i^z
	+
	\sum_{i<j}J_{ij}\tau_i^z\tau_j^z
	+
	\sum_{i<j<k}K_{ijk}\tau_i^z\tau_j^z\tau_k^z
	+\cdots,
\end{equation}
with coefficients decaying exponentially with the spatial extent of the operator support.
Thus, Floquet MBL, like static MBL, admits a complete quasi-local
commuting LIOM algebra. Consequently, the symmetry
criteria developed in the main text can be applied to the Floquet LIOM
algebra in the same way as for static MBL. For example, a Floquet MBL
system with particle-number conservation belongs to the same
A$+$U(1) class as its static counterpart, while other onsite symmetries
similarly lead to the corresponding symmetry classes listed in
Table~I.

Periodically driven systems may also possess additional symmetries that
combine spatial operations with time translation. Two important
examples are time-glide and time-screw symmetries. A time-glide symmetry
combines a spatial reflection with a fractional time translation
$\delta t$, for example $\delta t=pT/q$ with
$p,q\in\mathbb{Z}$ and $0<p<q$, while a time-screw symmetry combines a
spatial translation or rotation with a time translation. Schematically,
the spatial part of such a symmetry relates the Hamiltonian at different
times according to
\begin{equation}\notag
	G H(t) G^{-1}=H(t+\delta t),
\end{equation}
where $G$ denotes the spatial operation and $\delta t$ specifies the
associated time translation. The precise form of this relation depends
on the particular space-time symmetry. Over one complete driving
period, it consequently induces a corresponding constraint on the
Floquet evolution.

These symmetries can act nontrivially on the Floquet LIOMs even when a
complete LIOM algebra exists. To illustrate the spatial
part of such an action, consider an operation that maps site $i$ to a
symmetry-related site $g(i)$. With a choice of LIOM basis compatible
with this spatial symmetry, its action may take the form
\begin{equation}\notag
	G\tau_i^zG^{-1}=\tau_{g(i)}^z,
\end{equation}
where $g(i)$ denotes the image of site $i$ under the spatial operation.
Thus, the symmetry need not leave each individual LIOM invariant;
instead, it may permute or otherwise relate different LIOMs within the
complete algebra. The time-translation component introduces an additional layer of
structure. Although Floquet LIOMs are time-independent operators once
a Floquet time frame is chosen, changing the stroboscopic origin of the
drive changes their representation through the corresponding
micromotion unitary, i.e., the time evolution operator over a single driving period. A space-time symmetry can therefore relate the
LIOM representation in one Floquet time frame to that in another.
Schematically, if $\tau_i^z(0)$ denotes a LIOM in a reference Floquet
time frame and $\tau_i^z(\delta t)$ denotes the corresponding LIOM in
the frame shifted by $\delta t$, the symmetry may impose a relation of
the form
\begin{equation}\notag
	G\,\tau_i^z(0)\,G^{-1}
	=
	\tau_{g(i)}^z(\delta t).
\end{equation}
Here, $\tau_{g(i)}^z(\delta t)$ represents the same emergent LIOM
structure in a different Floquet time frame, rather than a
time-dependent LIOM. The fractional time translation therefore
constrains not only the spatial organization of the LIOMs but also how
the LIOM algebra is represented relative to the micromotion over a
driving cycle.

Consequently, Floquet space-time symmetries constrain the LIOM algebra
as a whole, including the relations among LIOMs at symmetry-related
spatial locations and their representations in different Floquet time
frames. They may also impose additional restrictions on the Floquet
operator and its quasienergy spectrum, including, depending on the
symmetry and its representation, possible degeneracies or pairing
relations that have no direct static counterpart. 

A systematic classification of Floquet MBL phases incorporating these
space-time symmetries would therefore require extending the present
framework to include their corresponding actions on the Floquet LIOM
algebra. Such an extension is beyond the scope of this work.
Nevertheless, the central operator-algebraic principle remains
unchanged: once a complete quasi-local Floquet LIOM algebra exists, the
stability of the corresponding MBL phase can be assessed by determining
whether the symmetry enforces degeneracies.
This provides a natural starting point for extending the present
classification to Floquet MBL phases with dynamical space-time
symmetries.


\begin{thebibliography}{99}
 \bibitem{AL1} P. W. Anderson, Absence of diffusion in certain random lattices, \href{https://journals.aps.org/pr/abstract/10.1103/PhysRev.109.1492} {Phys. Rev. {\bf 109}, 1492 (1958)}.
 \bibitem{AL2} P. A. Lee and T. V. Ramakrishnan, Disordered electronic systems, \href{https://journals.aps.org/rmp/abstract/10.1103/RevModPhys.57.287} {Rev. Mod. Phys. {\bf 57}, 287 (1985)}.
 \bibitem{AL3} B. Kramer and A. MacKinnon, Localization: Theory and experiment, \href{https://iopscience.iop.org/article/10.1088/0034-4885/56/12/001} {Rep. Prog. Phys. {\bf 56}, 1469 (1993)}.
 \bibitem{AL4} F. Evers and A. D. Mirlin, Anderson transitions, \href{https://journals.aps.org/rmp/abstract/10.1103/RevModPhys.80.1355} {Rev. Mod. Phys. {\bf 80}, 1355 (2008)}.
 \bibitem{AL5} Y. Wang, Emergent Dynamical Translational Symmetry Breaking as an Order Principle for Localization and Topological Transitions, \href{https://arxiv.org/pdf/2511.04360} {arXiv:2511.04360}.
 \bibitem{AZ1997} A. Altland and M. R. Zirnbauer, Nonstandard symmetry classes in mesoscopic normal-superconducting hybrid structures, \href{https://journals.aps.org/prb/abstract/10.1103/PhysRevB.55.1142} {Phys. Rev. B {\bf 55}, 1142 (1997)}.
 \bibitem{AZ1996} M. R. Zirnbauer, Riemannian symmetric superspaces and their origin in random-matrix theory,  \href{https://pubs.aip.org/aip/jmp/article/37/10/4986/454609/Riemannian-symmetric-superspaces-and-their-origin} {J. Math. Phys. {\bf 37}, 4986 (1996)}.
 \bibitem{ALS1} Y. Asada, K. Slevin, and T. Ohtsuki, Anderson Transition in
 Two-Dimensional Systems with Spin-Orbit Coupling, \href{https://journals.aps.org/prl/abstract/10.1103/PhysRevLett.89.256601}
 {Phys. Rev. Lett. {\bf 89}, 256601 (2002)}.
 \bibitem{ALS2} Y. Asada, K. Slevin, and T. Ohtsuki, Anderson transition in
 the three dimensional symplectic universality class, \href{https://journals.jps.jp/doi/10.1143/JPSJS.74S.238} {J. Phys.
 	Soc. Jpn. {\bf 74}, 238 (2005)}.
 \bibitem{ALS3}  S. Ryu, C. Mudry, H. Obuse, and A. Furusaki, $Z_2$ Topological Term, the Global Anomaly, and the
 Two-Dimensional Symplectic Symmetry Class of Anderson Localization, \href{https://journals.aps.org/prl/abstract/10.1103/PhysRevLett.99.116601} {Phys. Rev. Lett. {\bf 99}, 116601 (2007)}.
 \bibitem{ALS4}  A. D. Mirlin, F. Evers, I. V. Gornyi, and P.M. Ostrovsky,
 Anderson transitions: Criticality, symmetries, and topologies, \href{https://www.worldscientific.com/doi/abs/10.1142/S0217979210064526} {Int. J. Mod. Phys. B {\bf 24}, 1577 (2010)}.
 \bibitem{ALS5}  E. J. K\"{o}nig, P. M. Ostrovsky, I. V. Protopopov, and A. D. Mirlin, Metal-insulator transition in two-dimensional random fermion systems of chiral symmetry classes, \href{https://journals.aps.org/prb/abstract/10.1103/PhysRevB.85.195130} {Phys. Rev. B {\bf 85}, 195130 (2012)}.
 \bibitem{ALS6}  L. Fu and C. L. Kane, Topology, Delocalization via Average Symmetry and the Symplectic Anderson Transition,
 \href{https://journals.aps.org/prl/abstract/10.1103/PhysRevLett.109.246605} {Phys. Rev. Lett. {\bf 109}, 246605 (2012)}.
 \bibitem{ALS7} V. Kagalovsky, B. Horovitz, Y. Avishai, and J. T. Chalker,
 Quantum Hall Plateau Transitions in Disordered Superconductors, \href{https://journals.aps.org/prl/abstract/10.1103/PhysRevLett.82.3516} {Phys. Rev. Lett. {\bf 82}, 3516 (1999)}.
 \bibitem{ALS8} M. Ortu\~{o}, A. M. Somoza, and J. T. Chalker, Random
 Walks and Anderson Localization in a Three-Dimensional
 Class C Network Model, \href{https://journals.aps.org/prl/abstract/10.1103/PhysRevLett.102.070603} {Phys. Rev. Lett. {\bf 102}, 070603
 	(2009)}.
 \bibitem{ALS9} T. Wang, T. Ohtsuki, and R. Shindou, Universality classes of the
 Anderson transition in the three-dimensional symmetry classes
 AIII, BDI, C, D, and CI, \href{https://journals.aps.org/prb/abstract/10.1103/PhysRevB.104.014206} {Phys. Rev. B {\bf 104}, 014206 (2021)}.
 \bibitem{ALS10} X. Luo, Z. Xiao, K. Kawabata, T. Ohtsuki, and R. Shindou,
 Unifying the Anderson transitions in Hermitian and
 non-Hermitian systems, \href{https://journals.aps.org/prresearch/abstract/10.1103/PhysRevResearch.4.L022035} {Phys. Rev. Res. {\bf 4}, L022035 (2022)}.
 \bibitem{ALS11} Z. Xiao, K. Kawabata, X. Luo, T. Ohtsuki, and R. Shindou, Anisotropic Topological Anderson Transitions in Chiral Symmetry Classes, \href{https://journals.aps.org/prl/abstract/10.1103/PhysRevLett.131.056301} {Phys. Rev. Lett. {\bf 131}, 056301 (2023)}.
 
 \bibitem{TS1} A. P. Schnyder, S. Ryu, A. Furusaki, and A. W. W. Ludwig, Classification of topological insulators and superconductors in three spatial dimensions, \href{https://journals.aps.org/prb/abstract/10.1103/PhysRevB.78.195125} {Phys. Rev. B {\bf 78}, 195125 (2008)}.
 \bibitem{TS2} S. Ryu, A. P. Schnyder, A. Furusaki, and A. W. W. Ludwig, 
 Topological insulators and superconductors: tenfold way and dimensional hierarchy, \href{https://iopscience.iop.org/article/10.1088/1367-2630/12/6/065010} {New J. Phys. 12 065010 (2010)}.
 \bibitem{MBL0} D. M. Basko, I. L. Aleiner, and B. L. Altshuler, Metal-insulator transition in a weakly interacting many-electron system with localized single-particle states, \href{https://www.sciencedirect.com/science/article/pii/S0003491605002630?via%3Dihub} {Ann. Phys. {\bf 321},
 	1126 (2006)}.
 \bibitem{MBL1} D. A. Abanin, E. Altman, I. Bloch, and M. Serbyn, Colloquium: Many-body localization, thermalization, and entanglement, \href{https://journals.aps.org/rmp/abstract/10.1103/RevModPhys.91.021001} {Rev. Mod. Phys. {\bf 91}, 021001 (2019)}.
 \bibitem{MBL2} R. Nandkishore and D. A. Huse, Many-body localization and thermalization in quantum statistical mechanics, \href{https://www.annualreviews.org/content/journals/10.1146/annurev-conmatphys-031214-014726}
 {Annu. Rev. Condens. Matter Phys. {\bf 6}, 15 (2015)}.
 \bibitem{MBL3} P. Sierant, M. Lewenstein, A. Scardicchio, L. Vidmar, and J. Zakrzewski, Many-body localization in the age of classical computing, \href{https://iopscience.iop.org/article/10.1088/1361-6633/ad9756} {Rep. Prog. Phys. {\bf 88}, 026502 (2025).} 
 \bibitem{MBL4}  E. Altman and R. Vosk, Universal Dynamics and Renormalization in Many-Body-Localized Systems, \href{https://www.annualreviews.org/content/journals/10.1146/annurev-conmatphys-031214-014701} {Annu. Rev.
 	Condens. Matter Phys. {\bf 6}, 383 (2015)}.
 \bibitem{MBL5} E. Altman, Many-body localization and quantum thermalization, \href{https://www.nature.com/articles/s41567-018-0305-7} {Nat. Phys. {\bf 14}, 979 (2018)}.
 \bibitem{MBL6} M. Schreiber, S. S. Hodgman, P. Bordia, H. P. L\"{u}schen,
 M. H. Fischer, R. Vosk, E. Altman, U. Schneider, and I. Bloch, Observation of many-body localization of interacting fermions in a quasirandom optical lattice, \href{https://www.science.org/doi/10.1126/science.aaa7432} {Science {\bf 349}, 842 (2015)}.
 \bibitem{MBL7} J.-y. Choi, S. Hild, J. Zeiher, P. Schau$\beta$, A. RubioAbadal, T. Yefsah, V. Khemani, D. A. Huse, I. Bloch,
 and C. Gross, Exploring the many-body localization
 transition in two dimensions, \href{https://www.science.org/doi/10.1126/science.aaf8834} {Science {\bf 352}, 1547 (2016)}.
 \bibitem{MBL8} P. Bordia, H. L\"{u}schen, S. Scherg, S. Gopalakrishnan, M. Knap, U. Schneider, and I. Bloch, Probing
 Slow Relaxation and Many-Body Localization in Two
 Dimensional Quasiperiodic Systems, \href{https://journals.aps.org/prx/abstract/10.1103/PhysRevX.7.041047} {Phys. Rev. X {\bf 7}, 041047 (2017)}.
 \bibitem{MBL9} H. P. L\"{u}schen, P. Bordia, S. Scherg, F. Alet, E. Altman, U. Schneider, and I. Bloch,  Observation of Slow Dynamics near the Many-Body Localization Transition
 in One-Dimensional Quasiperiodic Systems, \href{https://journals.aps.org/prl/abstract/10.1103/PhysRevLett.119.260401} {Phys. Rev. Lett. {\bf 119}, 260401 (2017)}.
 \bibitem{MBL10} A. Lukin, M. Rispoli, R. Schittko, M. E. Tai, A. M. Kaufman, S. Choi, V. Khemani, J. L\'{e}onard, and M. Greiner, Probing entanglement in a many-body localized system, \href{https://www.science.org/doi/10.1126/science.aau0818}
 {Science {\bf 364}, 256 (2019)}.
 \bibitem{MBL11} M. Rispoli, A. Lukin, R. Schittko, S. Kim, M. E. Tai,
 J. L\'{e}onard, and M. Greiner, Quantum critical behaviour
 at the many-body localization transition, \href{https://www.nature.com/articles/s41586-019-1527-2} {Nature {\bf 573},
 	385 (2019)}.
 \bibitem{MBL12} J. L\'{e}onard, S. Kim, M. Rispoli, A. Lukin, R. Schittko, J. Kwan, E. Demler, D. Sels, and M. Greiner, Probing the onset of quantum avalanches in a many-body localized
 system, \href{https://www.nature.com/articles/s41567-022-01887-3} {Nat. Phys. {\bf 19}, 481 (2023)}.
 \bibitem{MBL13} J. Smith, A. Lee, P. Richerme, B. Neyenhuis, P. W. Hess, P. Hauke, M. Heyl, D. A. Huse, and C. Monroe, Many-body localization in a quantum simulator with
 programmable random disorder, \href{https://www.nature.com/articles/nphys3783} {Nat. Phys. {\bf 12}, 907
 	(2016)}.
 \bibitem{MBL14} S. Choi, J. Choi, R. Landig, G. Kucsko, H. Zhou, J. Isoya,
 F. Jelezko, S. Onoda, H. Sumiya, V. Khemani, C. von
 Keyserlingk, N. Y. Yao, E. Demler, and M. D. Lukin,
 Observation of discrete time-crystalline order in a disordered dipolar many-body system, \href{https://www.nature.com/articles/nature21426} {Nature {\bf 543}, 221 (2017)}.
 \bibitem{MBL15} P. Roushan, C. Neill, J. Tangpanitanon, V. M. Bastidas,
 A. Megrant, R. Barends, Y. Chen, Z. Chen, B. Chiaro,
 A. Dunsworth, A. Fowler, B. Foxen, M. Giustina, E. Jeffrey, J. Kelly, E. Lucero, J. Mutus, M. Neeley, C. Quintana, J. W. T. W. D. Sank, A. Vainsencher, H. Neven,
 D. G. Angelakis, and J. Martinis, Spectroscopic signatures of localization with interacting photons in superconducting qubits, \href{https://www.science.org/doi/10.1126/science.aao1401} {Science {\bf 358}, 1175 (2017)}.
 \bibitem{MBL16} K. Xu, J.-J. Chen, Y. Zeng, Y.-R. Zhang, C. Song,
 W. Liu, Q. Guo, P. Zhang, D. Xu, H. Deng, K. Huang,
 H. Wang, X. Zhu, D. Zheng, and H. Fan, Emulating
 Many-Body Localization with a Superconducting Quantum Processor, \href{https://journals.aps.org/prl/abstract/10.1103/PhysRevLett.120.050507} {Phys. Rev. Lett. {\bf 120}, 050507 (2018)}.
 \bibitem{MBL17} Q. Guo, C. Cheng, Z.-H. Sun, Z. Song, H. Li, Z. Wang,
 W. Ren, H. Dong, D. Zheng, Y.-R. Zhang, R. Mondaini,
 H. Fan, and H. Wang, Observation of energy-resolved
 many-body localization, \href{https://www.nature.com/articles/s41567-020-1035-1} {Nat. Phys. {\bf 17}, 234 (2019)}.
 \bibitem{MBL18} Y. Yao, L. Xiang, Z. Guo, Z. Bao, Y.-F. Yang, Z. Song,
 H. Shi, X. Zhu, F. Jin, J. Chen, S. Xu, Z. Zhu, F. Shen,
 N. Wang, C. Zhang, Y. Wu, Y. Zou, P. Zhang, H. Li,
 Z. Wang, C. Song, C. Cheng, R. Mondaini, H. Wang,
 J. Q. You, S.-Y. Zhu, L. Ying, and Q. Guo, Observation
 of many-body Fock space dynamics in two dimensions,
 \href{https://www.nature.com/articles/s41567-023-02133-0} {Nat. Phys. {\bf 19}, 1459 (2023)}.
 \bibitem{classA1} A. Pal and D. A. Huse, Many-body localization phase transition, \href{https://journals.aps.org/prb/abstract/10.1103/PhysRevB.82.174411} {Phys. Rev. B {\bf 82}, 174411 (2010)}.
 \bibitem{classA2} M. \v{Z}nidari\v{c}, T. Prosen, and P. Prelov\v{s}ek, Many-body localization in the Heisenberg magnet in a random field, \href{https://journals.aps.org/prb/abstract/10.1103/PhysRevB.77.064426} {Phys. Rev. B {\bf 77}, 064426 (2008)}.
 \bibitem{classA3} D. J. Luitz, N. Laflorencie, and F. Alet, Many-body localization edge in the random-field Heisenberg chain, \href{https://journals.aps.org/prb/abstract/10.1103/PhysRevB.91.081103} {Phys. Rev. B {\bf 91}, 081103(R) (2015)}.
 
 \bibitem{SU21} R. Vasseur, A. C. Potter, and S. A. Parameswaran, Quantum
 Criticality of Hot Random Spin Chains, \href{https://journals.aps.org/prl/abstract/10.1103/PhysRevLett.114.217201} {Phys. Rev. Lett. {\bf 114}, 217201 (2015)}.
 \bibitem{SU22} A. C. Potter and R. Vasseur, Symmetry constraints on many
 body localization, \href{https://journals.aps.org/prb/abstract/10.1103/PhysRevB.94.224206} {Phys.Rev.B {\bf 94}, 224206 (2016)}.
 \bibitem{SU23} I. V. Protopopov, W. W. Ho, and D. A. Abanin, Effect of
 SU(2) symmetry on many-body localization and thermalization, \href{https://journals.aps.org/prb/abstract/10.1103/PhysRevB.96.041122} {Phys. Rev. B {\bf 96}, 041122(R) (2017)}.
 \bibitem{SU24} I. V. Protopopov, R. K. Panda, T. Parolini, A. Scardicchio,
 E. Demler, and D. A. Abanin, Non-Abelian Symmetries and
 Disorder: A Broad Nonergodic Regime and Anomalous Thermalization, \href{https://journals.aps.org/prx/abstract/10.1103/PhysRevX.10.011025} {Phys.Rev.X {\bf 10}, 011025 (2020)}.
 \bibitem{SU25} S. Pai, N. S. Srivatsa, and A. E. B. Nielsen, Disordered Haldane-Shastry model, \href{https://journals.aps.org/prb/abstract/10.1103/PhysRevB.102.035117} {Phys. Rev. B {\bf 102}, 035117 (2020)}.
 \bibitem{SU26} P. Glorioso, L. V. Delacr\'{e}taz, X. Chen, R. M. Nandkishore, and
 A. Lucas, Hydrodynamics in lattice models with continuous
 non-Abelian symmetries, \href{https://scipost.org/10.21468/SciPostPhys.10.1.015} {SciPost Phys. {\bf 10}, 015 (2021)}.
 \bibitem{SU27} Z.-C. Yang, S. Nicholls, and M. Cheng, Extended nonergodic regime and spin subdiffusion in disordered SU(2)-symmetric Floquet systems, \href{https://journals.aps.org/prb/abstract/10.1103/PhysRevB.102.214205} {Phys. Rev. B {\bf 102}, 214205 (2020)}.
 \bibitem{SU28}  S A Parameswaran and R. Vasseur, Many-body localization, symmetry and topology, \href{https://iopscience.iop.org/article/10.1088/1361-6633/aac9ed} {Rep. Prog. Phys. {\bf 81}, 082501 (2018)}.
 \bibitem{SU29} D. Saraidaris, J.-W. Li, A. Weichselbaum, J. von Delft, and D. A. Abanin, Finite-size subthermal regime in disordered SU(N)-symmetric Heisenberg chains, \href{https://journals.aps.org/prb/abstract/10.1103/PhysRevB.109.094201} {Phys. Rev. B {\bf 109}, 094201 (2024)}.
 
 \bibitem{IOMclassA} M. Serbyn, Z. Papi\'{c}, and D. A. Abanin, Local Conservation Laws and the Structure of the Many-Body Localized States, \href{https://journals.aps.org/prl/abstract/10.1103/PhysRevLett.111.127201} {Phys. Rev. Lett. {\bf 111}, 127201 (2013)}.
 \bibitem{IOMSPT} D. A. Huse, R. Nandkishore, and V. Oganesyan, Phenomenology of fully many-body-localized systems, \href{https://journals.aps.org/prb/abstract/10.1103/PhysRevB.90.174202} {Phys. Rev. B {\bf 90}, 174202 (2014)}.
 \bibitem{SPT2} D. A. Huse, R. Nandkishore, V. Oganesyan, A. Pal, and S. L. Sondhi, Localization-protected quantum order, \href{https://journals.aps.org/prb/abstract/10.1103/PhysRevB.88.014206} {Phys. Rev. B {\bf 88}, 014206 (2013)}.
 \bibitem{SPT3} A. Chandran, V. Khemani, C. R. Laumann, and S. L. Sondhi, Many-body localization and symmetry-protected topological order, \href{https://journals.aps.org/prb/abstract/10.1103/PhysRevB.89.144201} {Phys. Rev. B {\bf 89}, 144201 (2014)}.
 \bibitem{SPT4} Y. Bahri, R. Vosk, E. Altman, and A. Vishwanath,  Localization and topology protected quantum coherence at the edge of hot matter, \href{https://www.nature.com/articles/ncomms8341} {Nat. Commun. {\bf 6}, 7341 (2015)}
 \bibitem{SPT5} A. Chan and T. B. Wahl, Classification of symmetry-protected topological many-body localized phases
 in one dimension, \href{https://iopscience.iop.org/article/10.1088/1361-648X/ab7f01} {J. Phys.: Condens. Matter. {\bf 32}, 305601 (2020).}
 \bibitem{SPT6} J. Li, A. Chan, and T. B. Wahl, Classification of symmetry-protected topological phases
 in two-dimensional many-body localized systems, \href{https://journals.aps.org/prb/abstract/10.1103/PhysRevB.102.014205} {Phys. Rev. B {\bf 102}, 014205 (2020)}.
 \bibitem{SPTMBL2} T. B. Wahl, Tensor networks demonstrate the robustness of localization and symmetry-protected topological phases, \href{https://journals.aps.org/prb/abstract/10.1103/PhysRevB.98.054204} {Phys. Rev. B {\bf 98}, 054204 (2018)}.
 \bibitem{SM} See the Supplemental Material for details on: (I) the Schrieffer--Wolff
 	construction of LIOMs, including representative disordered spin and
 	fermion models and the role of U(1) symmetry in their LIOM algebras;
 	(II) the derivation of Eq.~(4); (III) a possible physical interpretation of fragile MBL in class AII; and
 	(IV) the applicability of the LIOM classification to Stark MBL and
 	Floquet MBL, including additional Floquet space-time symmetries. The Supplemental Materials includes
 	the references \cite{StarkMBL4,FloquetMBL1}.
 \bibitem{classAI1} D. Pekker, G. Refael, E. Altman, E. Demler, and V. Oganesyan, Hilbert-Glass Transition: New Universality of Temperature-Tuned Many-Body Dynamical Quantum Criticality, \href{https://journals.aps.org/prx/abstract/10.1103/PhysRevX.4.011052} {Phys. Rev. X {\bf 4}, 011052 (2014)}.
 
 \bibitem{Dclass}  G. Kells, N. Moran, and D. Meidan, Localization enhanced and degraded topological order in interacting p-wave wires, \href{https://journals.aps.org/prb/abstract/10.1103/PhysRevB.97.085425} {Phys. Rev. B {\bf 97}, 085425 (2018)}.
 
 
 \bibitem{StarkMBL1} M. Schulz, C. A. Hooley, R. Moessner, and F. Pollmann, Stark Many-Body Localization, \href{https://doi.org/10.1103/PhysRevLett.122.040606} {Phys. Rev. Lett. {\bf 122}, 040606 (2019).}
 \bibitem{StarkMBL2} E. van Nieuwenburg, Y. Baum, and G. Refael, From Bloch Oscillations to Many Body Localization in Clean Interacting Systems, \href{https://www.pnas.org/doi/full/10.1073/pnas.1819316116} {Proc. Natl Acad. Sci. {\bf 116}, 9269 (2019)}.
 \bibitem{StarkMBL3} W. Morong, F. Liu, P. Becker, K. S. Collins, L. Feng, A. Kyprianidis, G. Pagano, T. You, A. V. Gorshkov, and C. Monroe, Observation of Stark many-body localization without disorder, \href{https://www.nature.com/articles/s41586-021-03988-0} {Nature {\bf 599}, 393 (2021).}
 \bibitem{StarkMBL4} T. M. Gunawardana and B. Bu\v{c}a, Dynamical l-bits and persistent oscillations in Stark many-body localization, \href{https://journals.aps.org/prb/abstract/10.1103/PhysRevB.106.L161111} {Phys. Rev. B {\bf 106}, L161111 (2022)}.
 \bibitem{FloquetMBL} P. Bordia, H. L\"{u}schen, U. Schneider, M. Knap, and I. Bloch, Periodically driving a many-body localized quantum system, \href{https://www.nature.com/articles/nphys4020}{Nat. Phys. {\bf 13}, 460 (2017).}
 \bibitem{FloquetMBL1} P. Ponte, Z. Papi\'{c}, F. Huveneers, and D. A. Abanin, Many-Body Localization in Periodically Driven Systems, \href{https://journals.aps.org/prl/abstract/10.1103/PhysRevLett.114.140401} {Phys. Rev. Lett. {\bf 114}, 140401 (2015).}
 \bibitem{2DMBL1} P. Bordia, H. L\"{u}schen, S. Scherg, S. Gopalakrishnan, M. Knap, U. Schneider, and I. Bloch, Probing Slow Relaxation and Many-Body Localization in Two-Dimensional Quasiperiodic Systems, \href{https://journals.aps.org/prx/abstract/10.1103/PhysRevX.7.041047} {Phys. Rev. X {\bf 7}, 041047 (2017).}
 \bibitem{2DMBL2} T.-M. Li, et al., Many-body delocalization with a two-dimensional 70-qubit superconducting quantum simulator, \href{https://arxiv.org/pdf/2507.16882}{arXiv:2507.16882.} 
 \bibitem{EndMBL} M. G. Sousa, R. F. P. Costa, G. D. de Moraes Neto, and E. Vernek, From ergodicity to Stark many-body localization in spin chains with single-ion anisotropy, \href{https://arxiv.org/abs/2401.03111} {arXiv:2401.03111}.
 
 \end{thebibliography}

\begin{thebibliography}{99}
	\bibitem{DynamicalLbits} T. M. Gunawardana and B. Bu\v{c}a, Dynamical l-bits and persistent oscillations in Stark many-body localization, \href{https://journals.aps.org/prb/abstract/10.1103/PhysRevB.106.L161111} {Phys. Rev. B {\bf 106}, L161111 (2022)}.
	\bibitem{FloquetMBL1s} P. Ponte, Z. Papi\'{c}, F. Huveneers, and D. A. Abanin, Many-Body Localization in Periodically Driven Systems, \href{https://journals.aps.org/prl/abstract/10.1103/PhysRevLett.114.140401} {Phys. Rev. Lett. {\bf 114}, 140401 (2015).}
\end{thebibliography}
\end{document}